\documentclass[12pt]{article}
\pdfoutput=1
\usepackage{latexsym, graphicx} 
\usepackage{amsmath}
\usepackage{amssymb}
\usepackage{caption}
\usepackage{subcaption}
\usepackage{cite}
\usepackage[utf8]{inputenc}
\usepackage{booktabs}
\usepackage{float}

\newcommand{\arXiv}[1]{\href{http://www.arXiv.org/abs/#1}{arXiv:#1}}
\usepackage[colorlinks=true, linkcolor=blue, bookmarks=true]{hyperref}

\makeatletter
\renewcommand\section{\@startsection {section}{1}{\z@}%
	{-3.5ex \@plus -1ex \@minus -.2ex}
	{2.3ex \@plus.2ex}%
	{\normalfont\large\bfseries}}
\renewcommand\subsection{\@startsection{subsection}{2}{\z@}%
	{-3.25ex\@plus -1ex \@minus -.2ex}%
	{1.5ex \@plus .2ex}%
	{\normalfont\bfseries}}
\makeatother

\allowdisplaybreaks

\begin{document}	

\begin{titlepage}
	\vfill
	\begin{center}
		{\LARGE \bf A Graceful Exit for  the\vspace{6mm}\\ Cosmological Constant Damping Scenario}
		
		\vskip 20mm
		
		{\large  Oleg Evnin$^{a,b}$ and K\'evin Nguyen$^{b}$}
		
		\vskip 15mm
		
		$^a$ Department of Physics, Faculty of Science, Chulalongkorn University,\\
		Thanon Phayathai, Pathumwan, Bangkok 10330, Thailand \\
		\vskip 3mm
		$^b$ Theoretische Natuurkunde, Vrije Universiteit Brussel (VUB), and \\ International Solvay Institutes, Pleinlaan 2, B-1050 Brussels, Belgium\\
		
		\vskip 15mm
		
		{\small\noindent  {\tt oleg.evnin@gmail.com, Kevin.Huy.D.Nguyen@vub.be}}
		
	\end{center}
	\vfill

	\begin{center}
		{\bf ABSTRACT}
		\vspace{3mm}
	\end{center}
	
	We present a broad and simple class of scalar-tensor scenarios that successfully realize dynamical damping of the effective cosmological constant, therefore providing a viable dynamical solution to the fine-tuning or ``old'' cosmological constant problem. In contrast to early versions of this approach, pioneered in the works of A.~Dolgov in the 1980es, these do not suffer from unacceptable variations of Newton's constant, as one aims at a small but strictly positive (rather than zero) late-time curvature. In our approach, the original fine-tuning issue is traded for a hierarchy of couplings, and we further suggest a way to naturally generate this hierarchy based on fermion condensation and softly broken field shift symmetry.

	\vfill
	
\end{titlepage}

\section{Introduction}

The cosmological constant of the $\Lambda$CDM cosmological model is notorious for the discrepancy between theoretical predictions and its observed small value. Indeed, any type of vacuum energy density should contribute to the cosmological constant $\Lambda$ and an inventory of the known sources of vacuum energy directly leads to estimates that are off by many orders of magnitude. See \cite{Martin:2012bt} for a very detailed review. An illustration of this fact is given by the electroweak (EW) theory that provides the dominant contribution from the Standard Model of particle physics. In this case one expects a vacuum energy density of the order 
\begin{equation}
\label{rhoEW}
|\rho_{EW}|\approx 10^{8}\ \text{GeV}^4.
\end{equation}
This has a classical contribution coming from the value of the Higgs potential in the spontaneously broken phase and a purely quantum contribution coming from vacuum fluctuations of the Higgs and other fields \cite{Martin:2012bt}.
Without an incredible fine-tuning of the various contributions from all sectors of particle physics, it seems impossible to explain the observed value of the cosmological vacuum energy density,
\begin{equation}
\label{rhoLCDM}
\rho_{\Lambda}|_{obs}=\frac{\Lambda}{8\pi G}\Big|_{obs}\approx 10^{-47}\ \text{GeV}^4.
\end{equation} 
The offset between \eqref{rhoEW} and \eqref{rhoLCDM} is of 55 orders of magnitude, and is generically expected to be even more severe in theories beyond the Standard Model such as Grand Unified Theories (GUT).\\

Possible solutions to this unacceptable discrepancy have been investigated in many different theoretical frameworks, including supersymmetry, quantum cosmology or string theory and extra dimensions. In addition to the ``old'' cosmological constant problem exposed above, one should also mention the seemingly unrelated \textit{coincidence problem} stating that the cosmological constant and baryonic matter contributions to the total energy content of our present Universe are roughly of the same order. According to the standard $\Lambda$CDM model predicting that the cosmological constant will eventually largely dominate over matter, we hence seem to live in a very particular time. A larger part of the cosmological constant literature, including quintessence models, is actually concerned with the resolution of this second puzzle, sometimes referred to as the ``new'' cosmological constant problem. In this context, the old fine-tuning problem is \textit{assumed} to be solved in some way or another. For a review of the models mentioned here, see \cite{Weinberg:1988cp,Copeland:2006wr,Martin:2012bt} and references therein.\\

A majority of attempted solutions to the ``old'' cosmological constant share the attitude that one must look for a fundamental (symmetry) principle that regulates the value of the cosmological constant appearing in the microscopic Lagrangian of the theory. This proves very challenging. In this work, we shall concentrate on an attractive but rather different perspective pioneered by Dolgov \cite{Dolgov} in 1980es. In this approach, one relies on \textit{dynamical} relaxation or damping mechanisms for the effective cosmological constant, i.e. the Hubble rate (which is what one measures in practice). The late-time cosmological dynamics is then a slow expansion despite the fundamental cosmological constant may be large, which is typically due to specific nonminimal couplings to extra fields. The problem is thus addressed in terms of properties of dynamical solutions, rather than in terms of symmetry properties of the fundamental Lagrangian. Such dynamical mechanisms are very appealing as they could occur generically in effective low-energy theories. Compared to microscopic approaches mentioned in the previous paragraph such as supersymmetry, they are also much more economical as they work independently of fine details of possible UV completions of the Standard Model. Such scenarios have been considered in \cite{Dolgov,Ford:1987de,Suen:1988nb,Dolgov:1982qq,Glavan:2015ora,Dolgov:1996zg,Rubakov:1999bw,Emelyanov:2011wn} (see also \cite{Barr:1986ya,Barr:2006mp,Dolgov:2008rf,Emelyanov:2011kn,Shlaer:2017jkx} for additional related works). In these models, the spacetime late-time attractor solution features a complete screening of the bare cosmological constant $\Lambda$, and therefore asymptotes to Minkowski spacetime. However, the simplest realizations of this approach suffer from unrealistic modifications of gravity: either the effective Newton constant vanishes \cite{Dolgov,Ford:1987de,Suen:1988nb,Dolgov:1982qq,Glavan:2015ora} or gravitational waves are of longitudinal and vector types \cite{Dolgov:1996zg,Rubakov:1999bw}. Solutions that bypass these problems do exist \cite{Emelyanov:2011wn}, although they rely on very fine-tuned and arguably unnatural Lagrangian interactions. If these obstacles could be overcome, this class of models would provide a promising opportunity to solve the ``old'' cosmological constant problem within a low-energy effective field theory (EFT) framework. According to Sahni and Starobinsky, ``this aesthetically attractive possibility should be investigated further"\cite{Sahni:1999gb}.\\   

For completeness, we mention that a different type of models referred to as \textit{Fab Four} have been put forward more recently. These form the most general subclass of Horndeski scalar-tensor theories admitting flat Minkowski
solutions \cite{Charmousis:2011bf}. In Fab Four models, it is therefore possible to get a complete screening of the bare cosmological constant $\Lambda$ appearing in the gravitational action. Higher-derivative modifications of these, sometimes called \textit{Fab Five} 
models, possess de Sitter solutions whose Hubble rate does not necessarily satisfy the standard relation to the cosmological constant, $3H^2\neq \Lambda$ \cite{Saridakis:2010mf,Appleby:2012rx,Starobinsky:2016kua}. These, however, suffer from vanishing of the gravitational strength. 
Finally, a phenomenologically interesting class of Horndeski theories has been studied recently in \cite{Appleby:2018yci}. They possess the desirable feature of not screening matter and radiation and do not seem to suffer from modified gravity laws, leaving open a possibility to describe realistic cosmologies. The Lagrangian is however so specific that it can be seen as a form of fine-tuning.\\

The old literature on the cosmological constant damping scenario was typically concerned with approaching power-law expansion at late times (i.e., zero effective cosmological constant) starting with arbitrary values of the cosmological constant term in the fundamental Lagrangian. This usually involved potentials unbounded from below and run-away behaviors for scalar fields (`unstable fields'), which in turn was the main cause of the complete late-time decay of Newton's constant, a phenomenologically unacceptable feature of the models. From a contemporary perspective, it is however more natural to aim at small positive values of the late-time effective cosmological constant. This opens a simple pathway for stabilizing the run-away behavior of the scalar field by adding a positive term to the potential, while ending up with a late-time solution characterized by un-varying effective cosmological and Newton constants with phenomenologically attractive values. Here, we wish to construct a model of dynamical damping of the cosmological constant along these lines that would satisfy the following requirements:
\begin{itemize}
	\item It has a late-time de Sitter \textit{attractor} solution whose Hubble rate is much smaller than the one expected from the bare cosmological constant: $3H^2\ll \Lambda$,
	\item It does not feature modifications of General Relativity once the late-time de Sitter solution is reached,
	\item A suitable notion of naturalness from the perspective of EFT is incorporated with respect to the field couplings introduced.
\end{itemize}
Note that by \textit{attractor} solution we mean a late-time solution that is reached independently of initial conditions. Although we will show numerical evidence of such attractor behaviors, we will not try to prove it with mathematical rigor. This is left for future work. We emphasize that a model satisfying the above criteria has not yet appeared in the literature and the goal of this paper is to fill this gap. In addition to these very reasonable and desirable features, the proposed model is of great simplicity.\\

In Section \ref{Section:Model}, we present a class of scalar-tensor models and discuss the requirements that their solutions should meet in order to produce a satisfactory resolution of the old cosmological constant problem, without introducing significant deviations from General Relativity at late times. We proceed in Section \ref{Section:Attractor} with a study of parameter space regions that possess late-time solutions with attractor behaviors of the type discussed above. Small values of specific field couplings are still needed in order to generate a small late-time effective cosmological constant. However, already at this stage, there is an arguable gain: the \textit{fine-tuning} issue associated with the conventional view of the cosmological constant problem, i.e. the need for extremely precise cancellation among multiple large contributions, is traded for a \textit{hierarchy} among couplings. In section \ref{Section:Naturalness}, we suggest a way to generate the necessary small couplings, which relies on fermionic condensates and naturally produces huge ratios of scales, in the spirit of Quantum Chromodynamics (QCD). In such a formulation, one starts with a Lagrangian containing a chosen set of field couplings of order one in Planck units and an arbitrarily large bare cosmological constant $\Lambda$, while still ending up with a small Hubble rate at late times. We further comment on the naturalness of the coupling pattern introduced from an EFT perspective.


\section{The model} \label{Section:Model}

We consider a self-interacting scalar field nonminimally coupled to Einstein gravity with a positive cosmological constant,\footnote{See \cite{Glavan:2015aqa} for a similar action considered in a different context.}
\begin{align}\label{actiontotal}
S_{total}&=S_{EH}+S_\phi,\\
\label{EHaction}
S_{EH}&=\frac{1}{G} \int d^4x\ \sqrt{-g} \left(R-2\Lambda\right),\\
\label{phiaction}
S_\phi&=-\frac{1}{2} \int d^4x\ \sqrt{-g} \left[(\partial_\mu \phi)^2+m^2\phi^2+\xi R\phi^2+\lambda \phi^4+G\lambda_R R \phi^4\right].
\end{align}
Here, we have restricted the content of $S_\phi$ to polynomials\footnote{We note that a mention of quartic self-coupling of the scalar field appears already in \cite{Dolgov}. However, that paper makes an attempt, in accordance with the spirit of the time, to obtain a late-time solution with a vanishing effective cosmological constant, for which the quartic coupling is not helpful, and is hence dismissed. This is different from the contemporary perspective we adopt here.} of $\phi$ respecting the $\phi \to -\phi$ symmetry, renormalizable upon quantization of the scalar field in a classical background geometry \cite{Birrell:1982ix}. We have also restricted our analysis to terms depending at most linearly on the curvature, with nonminimal couplings $\xi$ and $\lambda_R$. A factor of $G$ appears in front of $R \phi^4$ so as to make $\lambda_R$ dimensionless. These terms linear in the curvature are necessary ingredients of the cosmological constant damping mechanism presented in this paper, while higher powers of the curvature appear naturally suppressed in Planck units and can be ignored. The cosmological constant $\Lambda$ is naturally related to the Hubble rate $H_\Lambda$ of the initial phase of exponential expansion through the equality $\Lambda\equiv 3H_\Lambda^2$.\\

We emphasize that the action (\ref{actiontotal}-\ref{phiaction}) is written in the Jordan frame, according to usual conventions in scalar-tensor theories (one notices the term $\xi R\phi^2$). As the definition of the cosmological constant term depends on the frame, it is important to keep in mind that we are solving the
cosmological constant in the Jordan frame. If converted to the Einstein frame, the Jordan frame cosmological constant term will look like a non-constant contribution to the scalar field potential, rather than like a cosmological constant term. Physically, one must solve the cosmological constant problem in the frame whose metric is seen by the matter fields. An attempt to consider matter fields coupled to the Einstein frame metric in a related context has been made in \cite{Glavan:2015ora}.\\
  
We shall assume homogeneity and isotropy of the cosmological solutions, so that one has to deal with a time-dependent scalar field $\phi(t)$ together with an FLRW metric ansatz,
\begin{equation}\label{metrFLRW}
ds^2=-dt^2+a(t)^2d\mathbf{x}^2, \qquad H(t)\equiv\frac{\dot{a}}{a}.
\end{equation}
The Klein-Gordon and Friedmann equations for the scalar field $\phi(t)$ and the Hubble rate $H(t)$ are 
\begin{align}
\label{KG}
\left[\partial_t^2+3H\partial_t +m^2+6\xi\left(\dot{H}+2H^2\right)\right]\phi&=-2\left[\lambda+6G\lambda_R \left(\dot{H}+2H^2\right)\right]\phi^3,\\
\label{Friedmann}
H^2&=\frac{G}{6} \rho+H_\Lambda^2,
\end{align}
where the scalar field energy density is
\begin{equation}
\label{rho}
\rho=\frac{1}{2}\left((\partial_t \phi)^2+m^2\phi^2+\lambda \phi^4\right)+\left(3H^2+3H\partial_t\right)\left(\xi\phi^2+G\lambda_R \phi^4\right).
\end{equation}
We want to consider a situation for which the cosmological constant $\Lambda$ appearing in the action \eqref{EHaction} is large compared to its observed value, $GH_\Lambda^2 \gg GH^2\big|_{obs} \sim 10^{-120}$. It could be of the Planck, GUT or EW symmetry breaking energy scale, for example. Such a situation would be free of fine-tuning referred to as the ``old'' cosmological constant problem. We will show that in appropriate regions of the coupling parameter space $(m,\xi,\lambda,\lambda_R)$, the system is \textit{dynamically attracted} towards very small constant Hubble rate,
\begin{equation}
\label{naivedamping}
\lim\limits_{t\to \infty}H(t)=H_0, \qquad  G H_0^2\ll G H_\Lambda^2.
\end{equation}
This will provide a dynamical mechanism for explaining the smallness of the observed \mbox{Hubble} rate and a possible resolution of the ``old'' cosmological constant problem.\\

Looking at the Friedmann equation \eqref{Friedmann}, it is obvious that the condition \eqref{naivedamping} can be reached dynamically only if the energy density grows to large negative values so as to compensate for the cosmological constant vacuum energy. As in the original works by Dolgov \cite{Dolgov} and Ford \cite{Ford:1987de}, the basic ingredient is to have a negative quadratic coupling to curvature, $\xi<0$. In such a case, the scalar field is unstable and $\phi^2$ grows to large values while $\rho$ becomes negative. However, as for any scalar-tensor gravitational theory, the presence of nonminimal couplings is hardly compatible with experiments as it results in a time-varying effective Newton constant $G_{eff}$, unless the scalar field is constant today. This is where most models of cosmological constant damping have failed. The role of the quartic couplings $\lambda$ and $\lambda_R$ considered in \eqref{phiaction}, which we assume to be positive, is to stabilize the scalar field to a constant value at late-time,
\begin{equation}
\lim\limits_{t\to \infty} \phi(t)=\phi_0.
\end{equation}
In this case, the late-time effective Newton constant is
\begin{equation}
\label{Geff}
\frac{1}{G_{eff}}=\frac{1}{G}-\frac{1}{2}\left( \xi \phi_0^2+G\lambda_R \phi_0^4\right).
\end{equation}
This also means that in the condition \eqref{naivedamping}, $G$ should be replaced on the left-hand side by $G_{eff}$ as this is the gravitational coupling measured at late-time once the attractor solution has been reached,
\begin{equation}
\label{damping}
G_{eff} H_0^2 \ll G H_\Lambda^2.
\end{equation}

One can contrast our approach more specifically with the early works of Dolgov \cite{Dolgov} and Ford \cite{Ford:1987de}, in which a strictly vanishing effective cosmological constant was aimed at. For this purpose, one sets $m^2=\lambda=\lambda_R=0$ and $\xi<0$ in the action \eqref{phiaction}, such that the model possesses an attractor solution whose late-time behavior is characterized by a linearly growing scalar field and power-law expansion,
\begin{equation}
\phi(t)\sim t, \qquad a(t)\sim t^{\alpha}, \qquad \alpha=\frac{2|\xi|+1}{4|\xi|}.
\end{equation}
Such a power-law dependence corresponds to a spacetime that is asymptotically Minkowski in the future, as can be manifested by introducing $\xi^i=a(t) x^i$ and rewriting the metric (\ref{metrFLRW}) in the coordinates $(t,\xi^i)$. This scenario dynamically solves the old cosmological constant problem,
\begin{equation}
G_{eff}H^2\sim t^{-4}, \qquad G_{eff}\sim t^{-2},
\end{equation}
however at the price of a vanishing effective gravitational strength, $G_{eff} \to 0$. Similar conclusions are reached in the massive case, $m^2>0$, as we show in the appendix. In fact, there is a well-known no-go theorem for obtaining such Minkowski attractor solutions  without turning off the effective gravitational coupling \cite{Weinberg:1988cp}. If one aims instead at a small but strictly positive curvature $H^2$, this no-go theorem does not apply, as our considerations explicitly demonstrate. We achieve this by turning on the quartic couplings $\lambda, \lambda_R>0$ that will stabilize the scalar field together with the effective Newton constant $G_{eff}$. 


\section{Attractor solutions} \label{Section:Attractor}

We shall now study various regions of the parameter space $(m^2,\xi,\lambda,\lambda_R)$ that possess \textit{attractor solutions} realizing the cosmological constant damping. We make the following assumptions:
\begin{itemize}
\item $\xi<0$ and $m^2, \lambda, \lambda_R, G, H_\Lambda >0$,
\item $Gm^2, \lambda, \lambda_R \ll |\xi|$ and $GH_\Lambda^2 < 1$.
\end{itemize}
We recall that the condition $\xi<0$ allows for growing/unstable solutions of $\phi$ and for the presence of a negative energy density component in \eqref{rho}. To be more precise, growing solutions arise in case that $m^2+\xi R <0$, with the curvature scalar $R=6\dot{H}+12H^2$. The condition $\lambda,\lambda_R >0$ ensures that the scalar potential is bounded from below, which means in particular that $\phi$ will eventually stabilize after an initial period of instability driven by the nonminimal coupling $\xi$. The assumption that several orders of magnitude separate $Gm^2,\lambda,\lambda_R$ from $\xi$ implies that stabilization will occur for large values of $\phi$, and hence for small values of $G_{eff}$, as can be seen from \eqref{Geff}. This turns out to be a crucial element in producing the Hubble rate damping \eqref{damping} that we are looking for. In addition, these simplifying assumptions are very natural from an effective low-energy perspective. Indeed, an initially large Hubble rate $GH_\Lambda^2$, originating from the fundamental cosmological constant of the theory, is driven towards its small observed value at late times by the infrared sector of the theory. We discuss the naturalness of these assumptions in more detail in Section \ref{Section:Naturalness}.\\
\renewcommand{\arraystretch}{1.8}
\begin{table}
	\begin{tabular}{l|ccccc}
		\toprule[0.05cm]
		& I & II & III & IV & V \\
		\midrule
		& {\small $Gm^2, \lambda_R \ll \lambda $} & {\small $Gm^2 \lesssim \lambda_R \sim \lambda $} & {\small $0<\lambda_R \lesssim \lambda \ll Gm^2$} & {\small $\lambda \ll \lambda_R \ll Gm^2$} & {\small $Gm^2,\lambda \ll \lambda_R$} \\
		\midrule[0.05cm]
		$G_{eff} H_0^2$ & \( \frac{\lambda}{3\xi^2} \) & \( \frac{\lambda}{3\xi^2} \) & \( \frac{\lambda}{3\xi^2} \) & \( \frac{9Gm^2\lambda_R}{4|\xi|^3} \) & \( \frac{64GH_\Lambda^2 \lambda_R^2}{\xi^4} \) \\
		$H_0^2$ & \( H_\Lambda^2 \) & \(  \left(\frac{H_\Lambda^2 \lambda^2}{18 G^2 \xi^2 \lambda_R}\right)^{1/3} \) & \(  \frac{1}{6}\sqrt{\frac{m^2\lambda}{2G|\xi|\lambda_R}} \) & \( \frac{m^2}{4|\xi|}\) & \( \frac{8H_\Lambda^2 \lambda_R}{\xi^2}\)\\
		$\phi_0^2$ & \( \frac{6|\xi|H_\Lambda^2}{\lambda}\) & \( \frac{6|\xi|H_0^2}{\lambda} \) & \( \sqrt{\frac{m^2 |\xi|}{2G\lambda\lambda_R}} \) & \( \frac{|\xi|}{3G\lambda_R}\)& \( \frac{|\xi|}{2G\lambda_R} \)\\
		$G_{eff}$ & \( \frac{\lambda}{3\xi^2 H_\Lambda^2} \) & \( \frac{\lambda}{3\xi^2H_0^2} \) & \( \sqrt{\frac{8G\lambda\lambda_R}{m^2 |\xi|}} \) & \( \frac{9G\lambda_R}{\xi^2} \) & \( \frac{8G\lambda_R}{\xi^2} \) \\
	\end{tabular}
    \caption{Characteristic quantities of late-time constant attractor solutions in relevant regions of the parameter space. Damping of the Hubble rate in effective Planck units, $G_{eff}H_0^2 \ll 1$, is achieved in all cases considered. 
    }
    \label{table}
\end{table}

\begin{figure}[h!]
	\centering
	\begin{subfigure}[b]{.45\textwidth}
		\centering
		\includegraphics[trim=3.5cm 8cm 4cm 8cm, clip=true, width=1\linewidth]{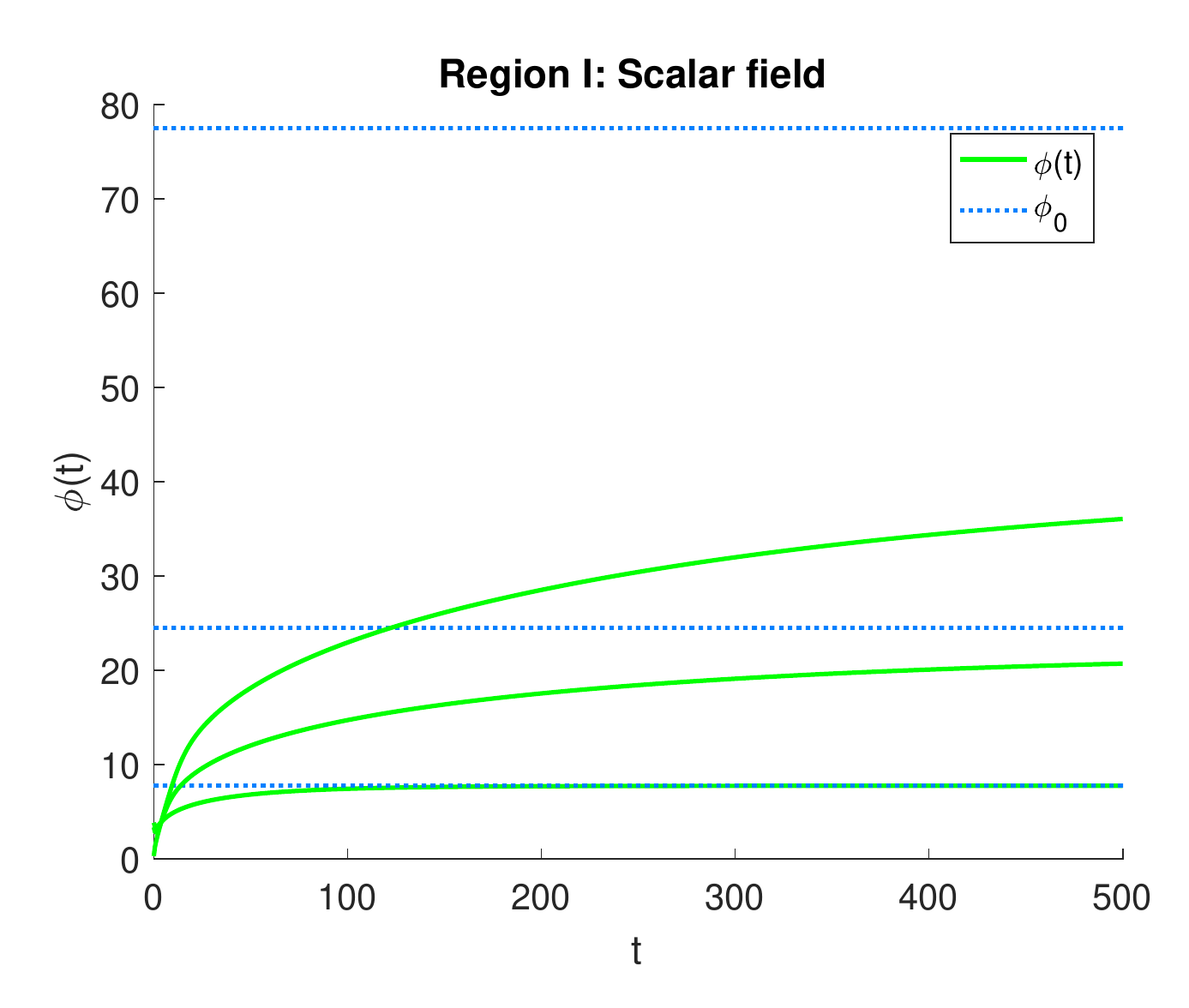}
		\caption{}
		\label{fig:regI:1}
	\end{subfigure}%
	\begin{subfigure}[b]{.45\textwidth}
		\centering
		\includegraphics[trim=3.5cm 8cm 4cm 8cm, clip=true, width=1\linewidth]{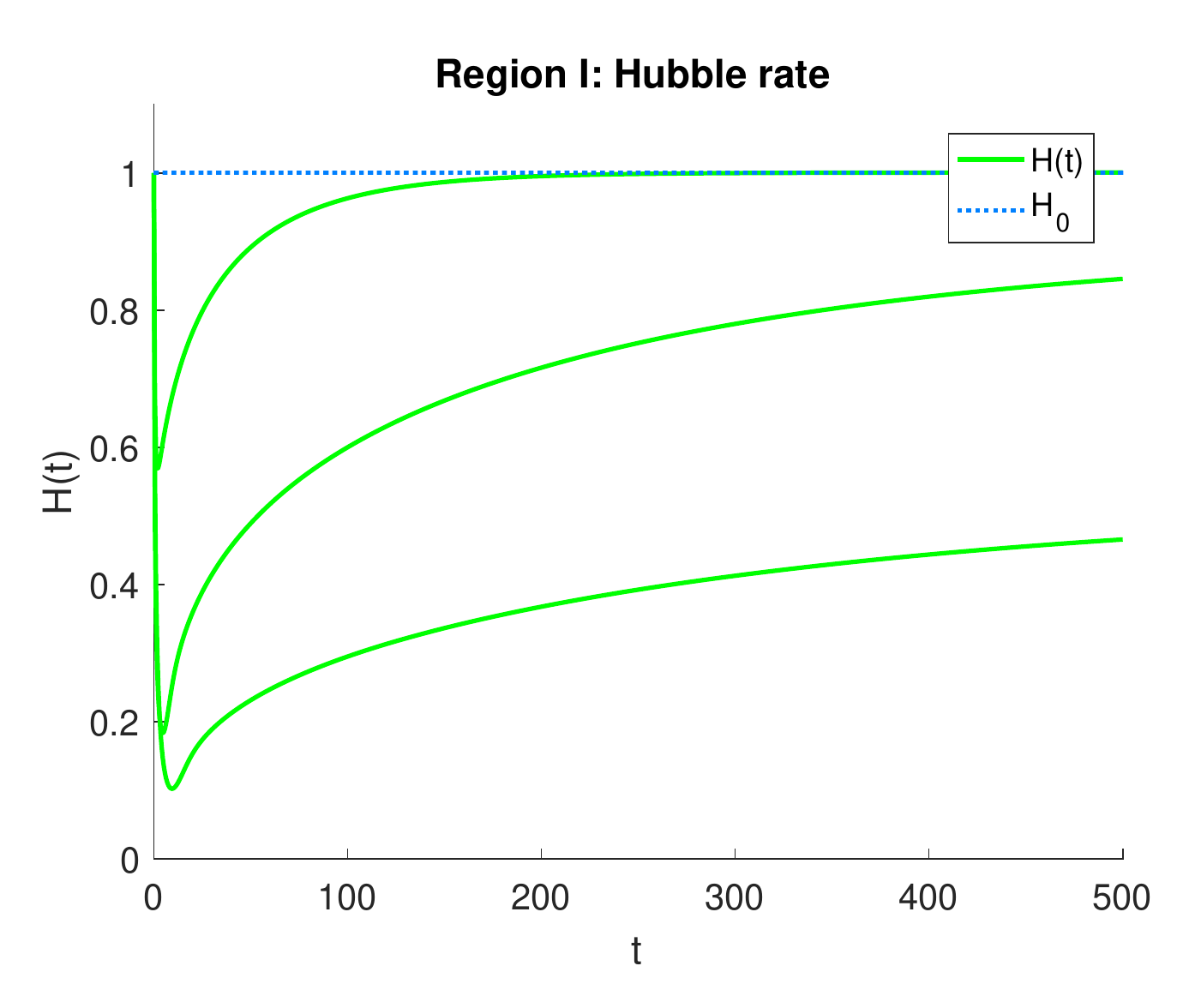}
		\caption{}
		\label{fig:regI:2}
	\end{subfigure}
	\begin{subfigure}[b]{.45\textwidth}
		\centering
		\includegraphics[trim=3.5cm 8cm 4cm 8cm, clip=true, width=1\linewidth]{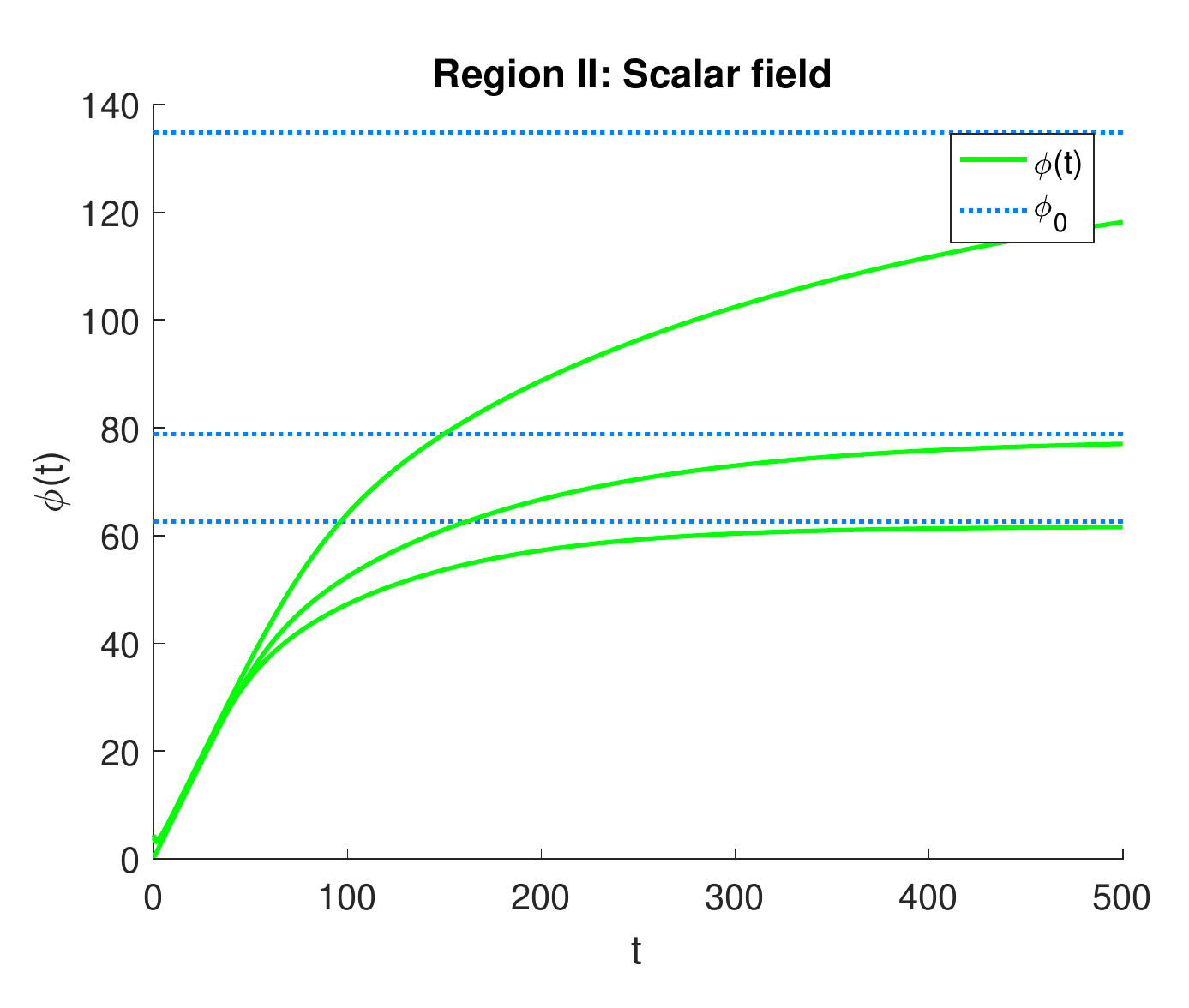}
		\caption{}
		\label{fig:regII:1}
	\end{subfigure}%
	\begin{subfigure}[b]{.45\textwidth}
		\centering
		\includegraphics[trim=3.5cm 8cm 4cm 8cm, clip=true, width=1\linewidth]{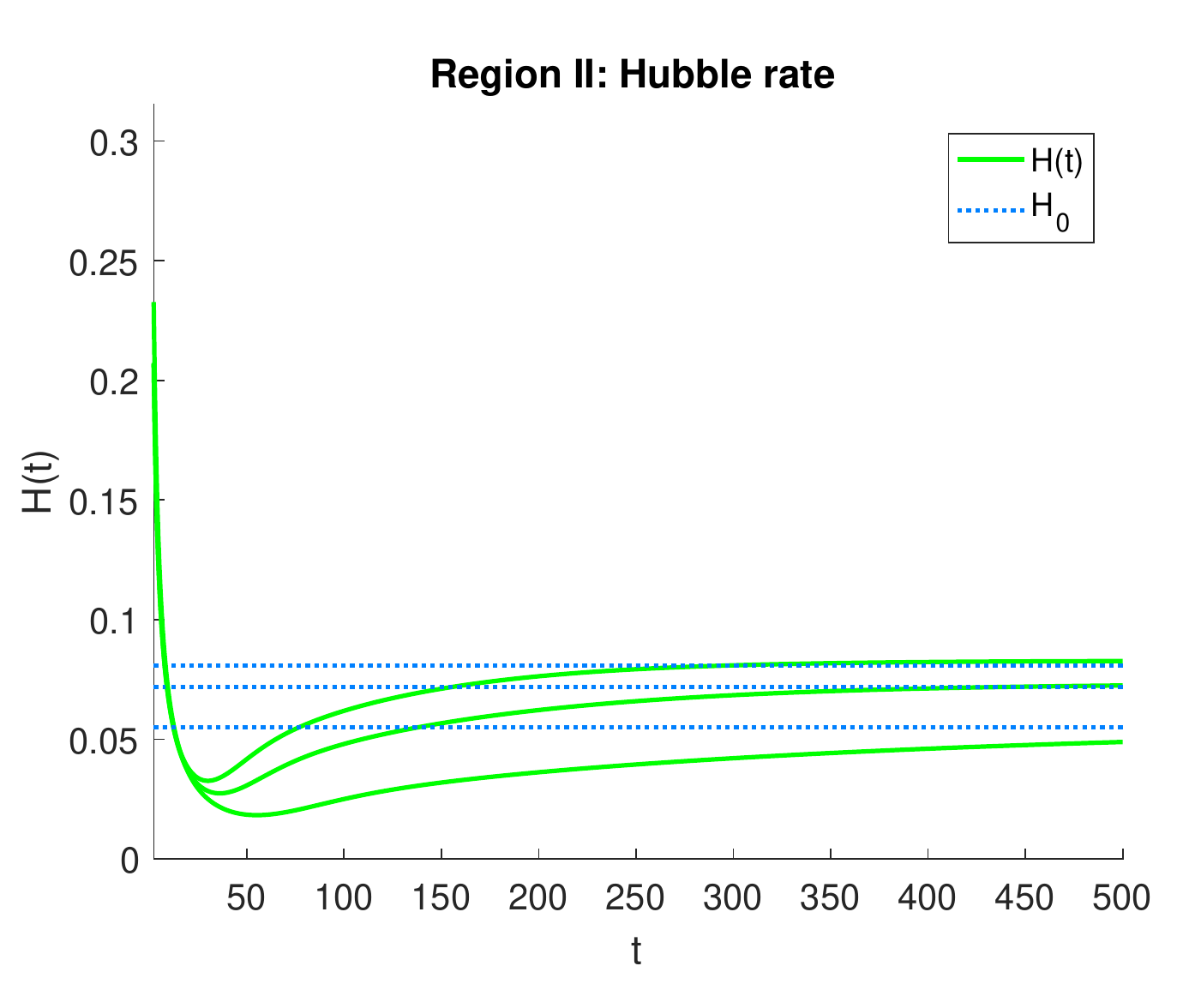}
		\caption{}
		\label{fig:regII:2}
	\end{subfigure}
	\caption{A sample of dynamical solutions, all of which display a constant late-time behavior, which does not depend on the choice of initial conditions. Figures \ref{fig:regI:1}-\ref{fig:regI:2}: Region I. Figures \ref{fig:regII:1}-\ref{fig:regII:2}: Region II.}
	\label{Fig:Region12}
\end{figure}
   
\begin{figure}[H]
	\centering
	\begin{subfigure}[b]{.45\textwidth}
		\centering
		\includegraphics[trim=3.5cm 8cm 4cm 8cm, clip=true, width=1\linewidth]{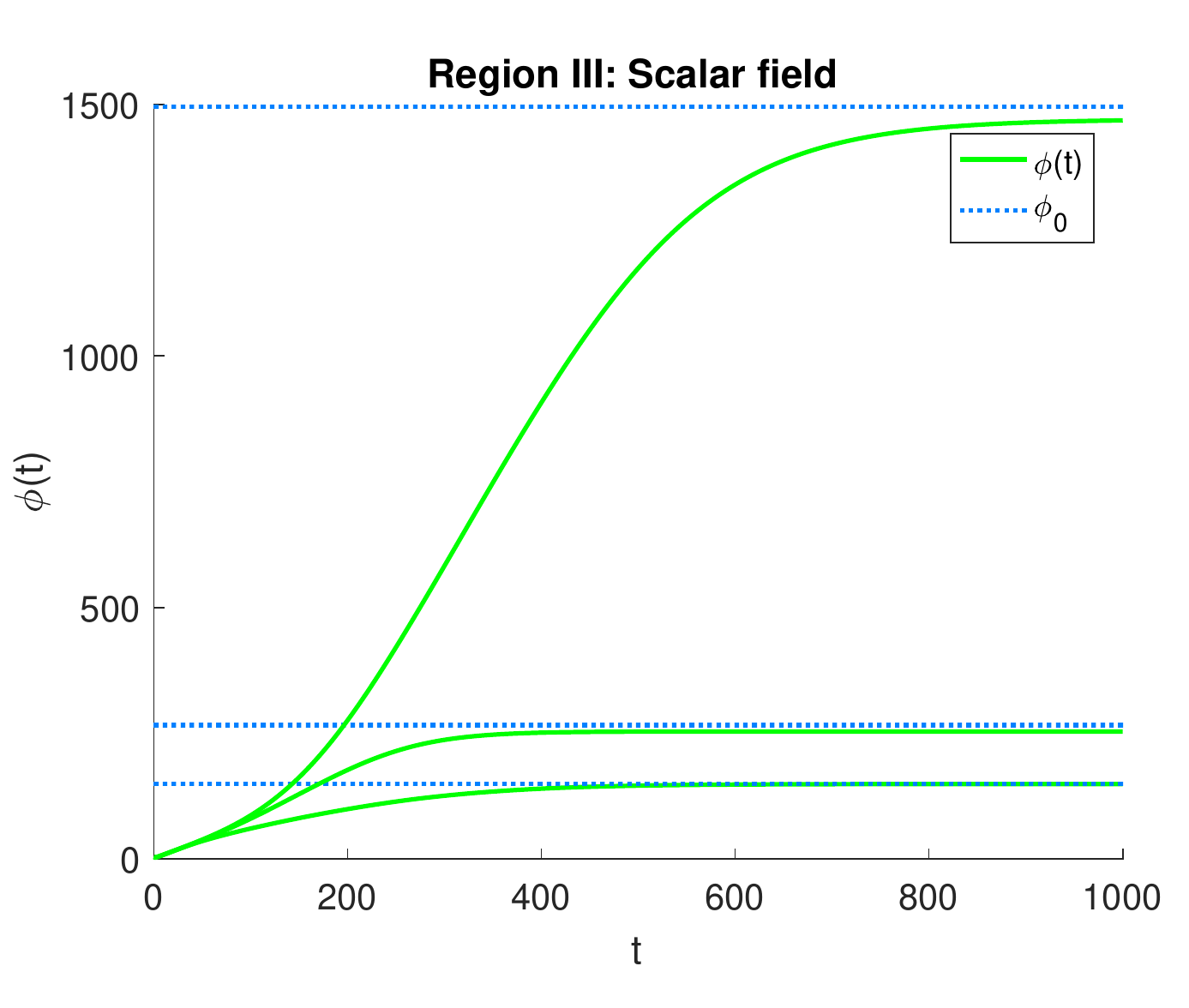}
		\caption{}
		\label{fig:regIII:1}
	\end{subfigure}%
	\begin{subfigure}[b]{.45\textwidth}
		\centering
		\includegraphics[trim=3.5cm 8cm 4cm 8cm, clip=true, width=1\linewidth]{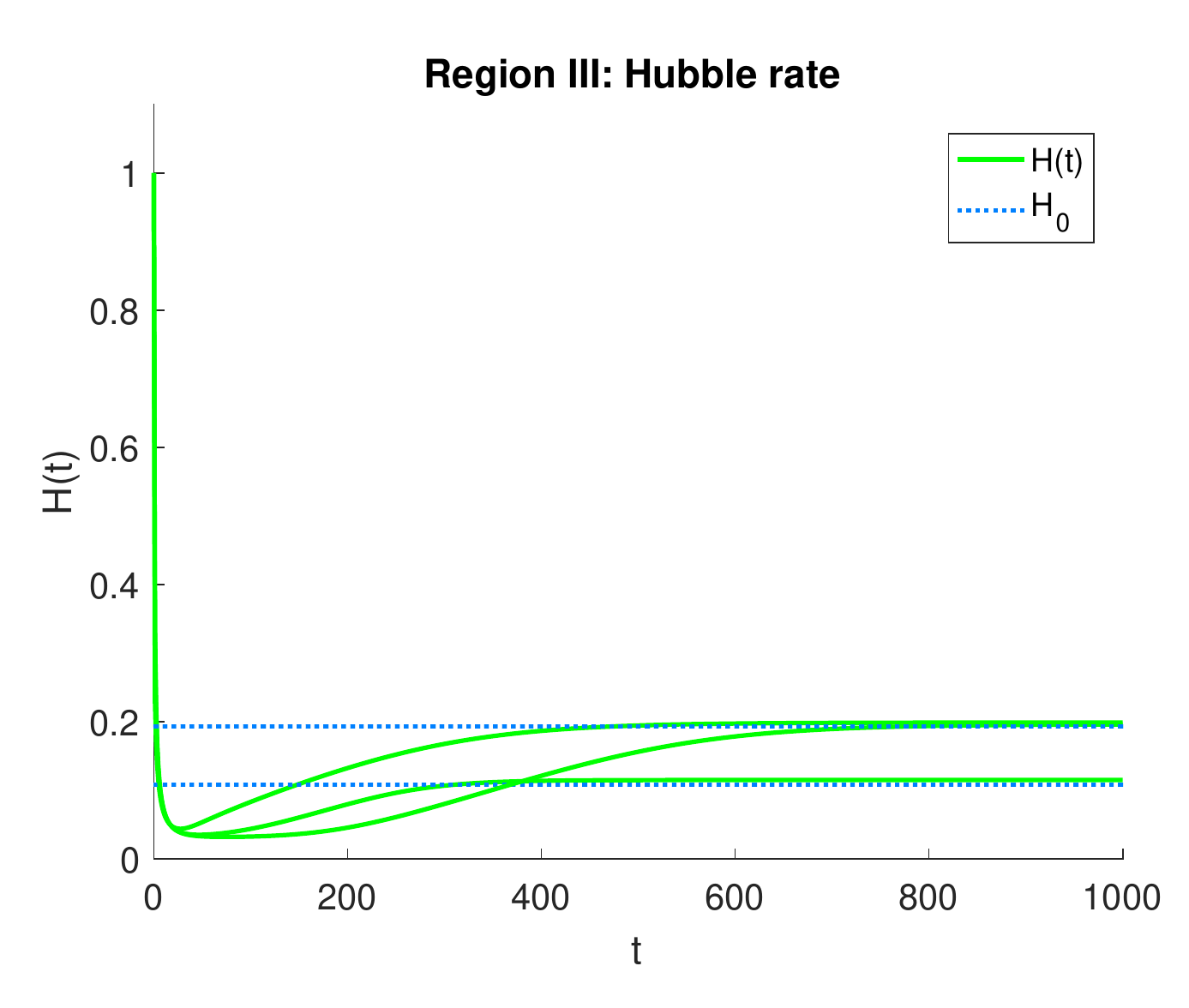}
		\caption{}
		\label{fig:regIII:2}
	\end{subfigure}
	\begin{subfigure}[b]{.33\textwidth}
		\centering
		\includegraphics[trim=3.5cm 8cm 4cm 8cm, clip=true, width=1\linewidth]{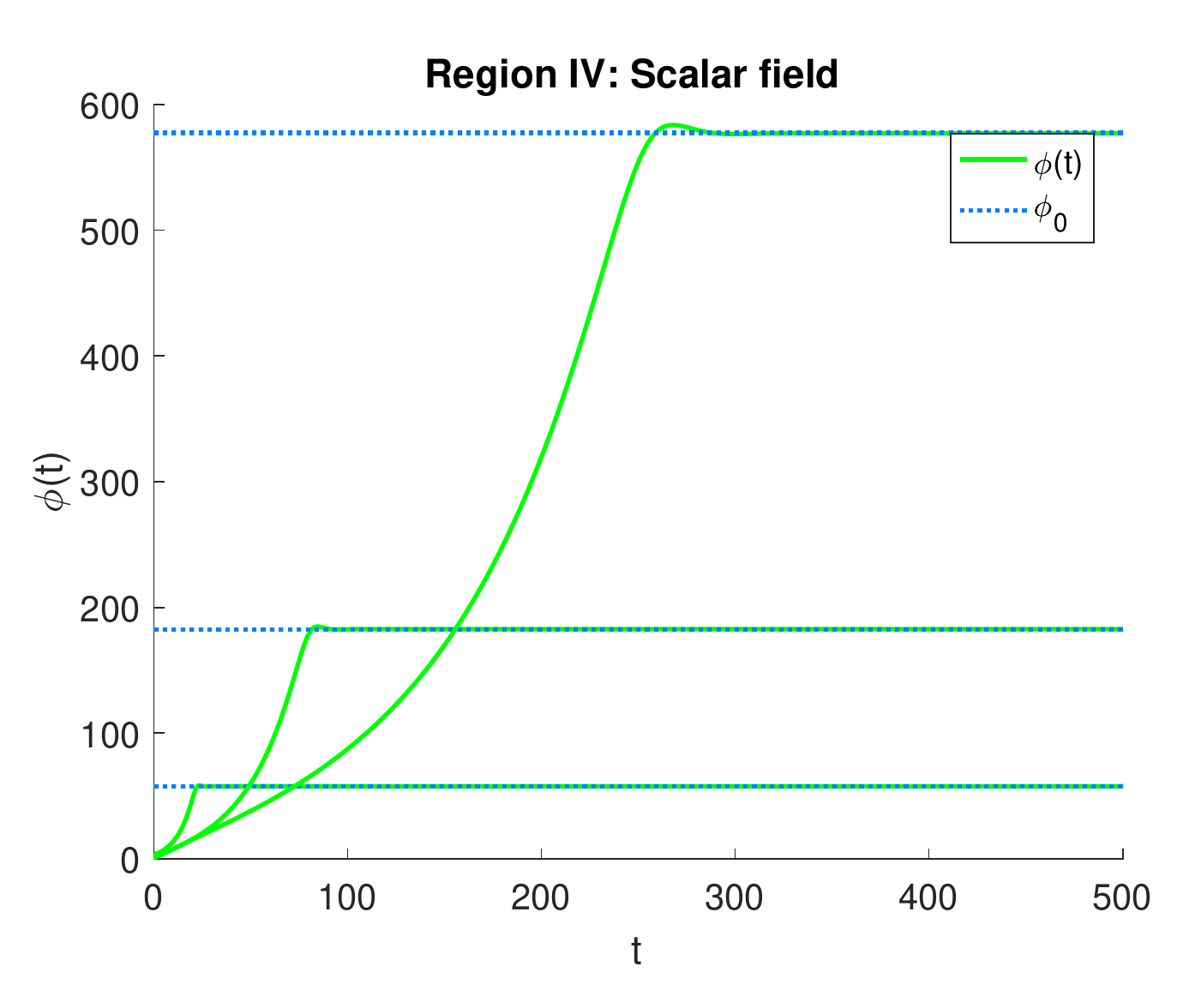}
		\caption{}
		\label{fig:regIV:1}
	\end{subfigure}%
	\begin{subfigure}[b]{.33\textwidth}
		\centering
		\includegraphics[trim=3.5cm 8cm 4cm 8cm, clip=true, width=1\linewidth]{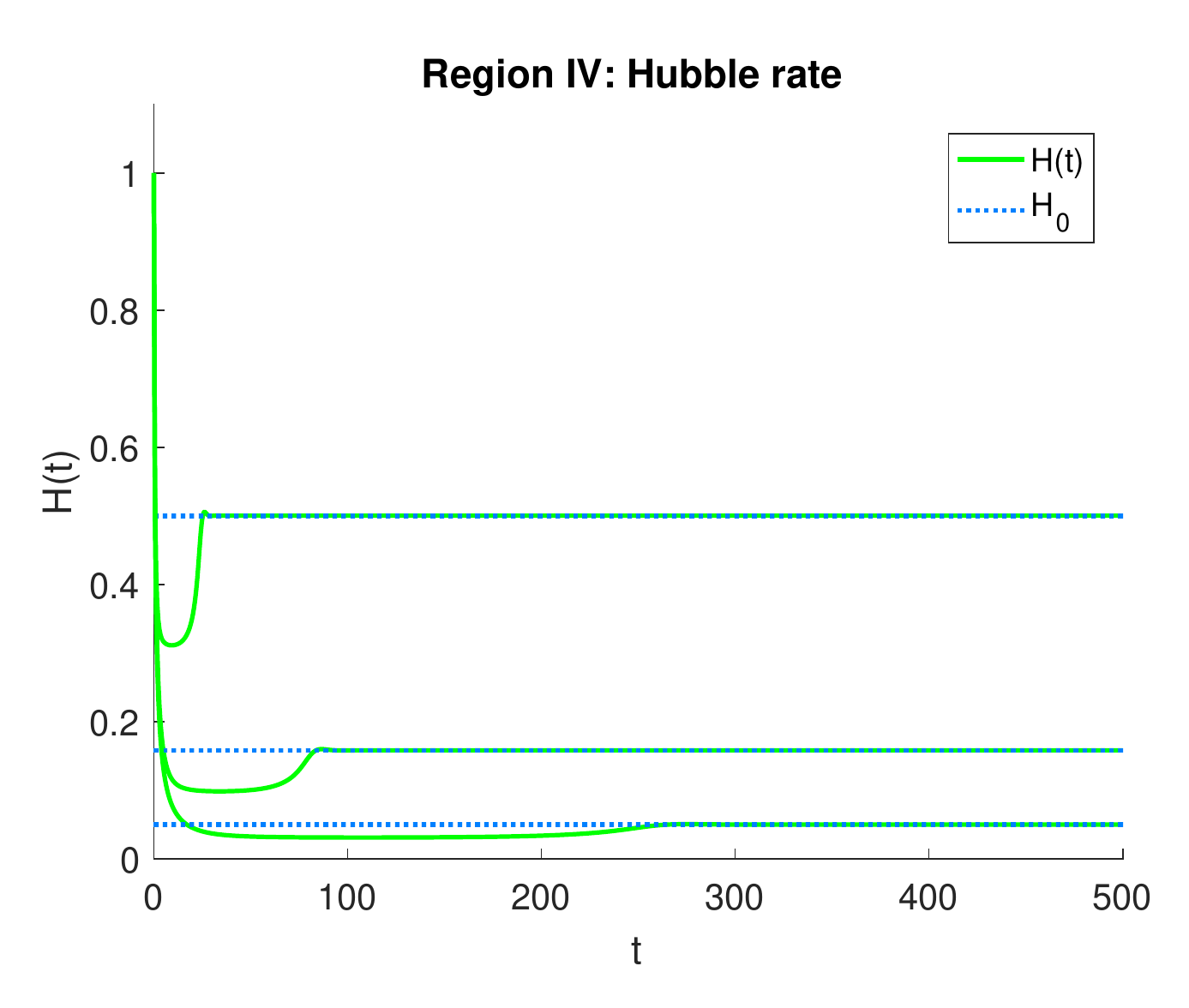}
		\caption{}
		\label{fig:regIV:2}
	\end{subfigure}
	\begin{subfigure}[b]{.33\textwidth}
		\centering
		\includegraphics[trim=3.5cm 8cm 4cm 8cm, clip=true, width=1\linewidth]{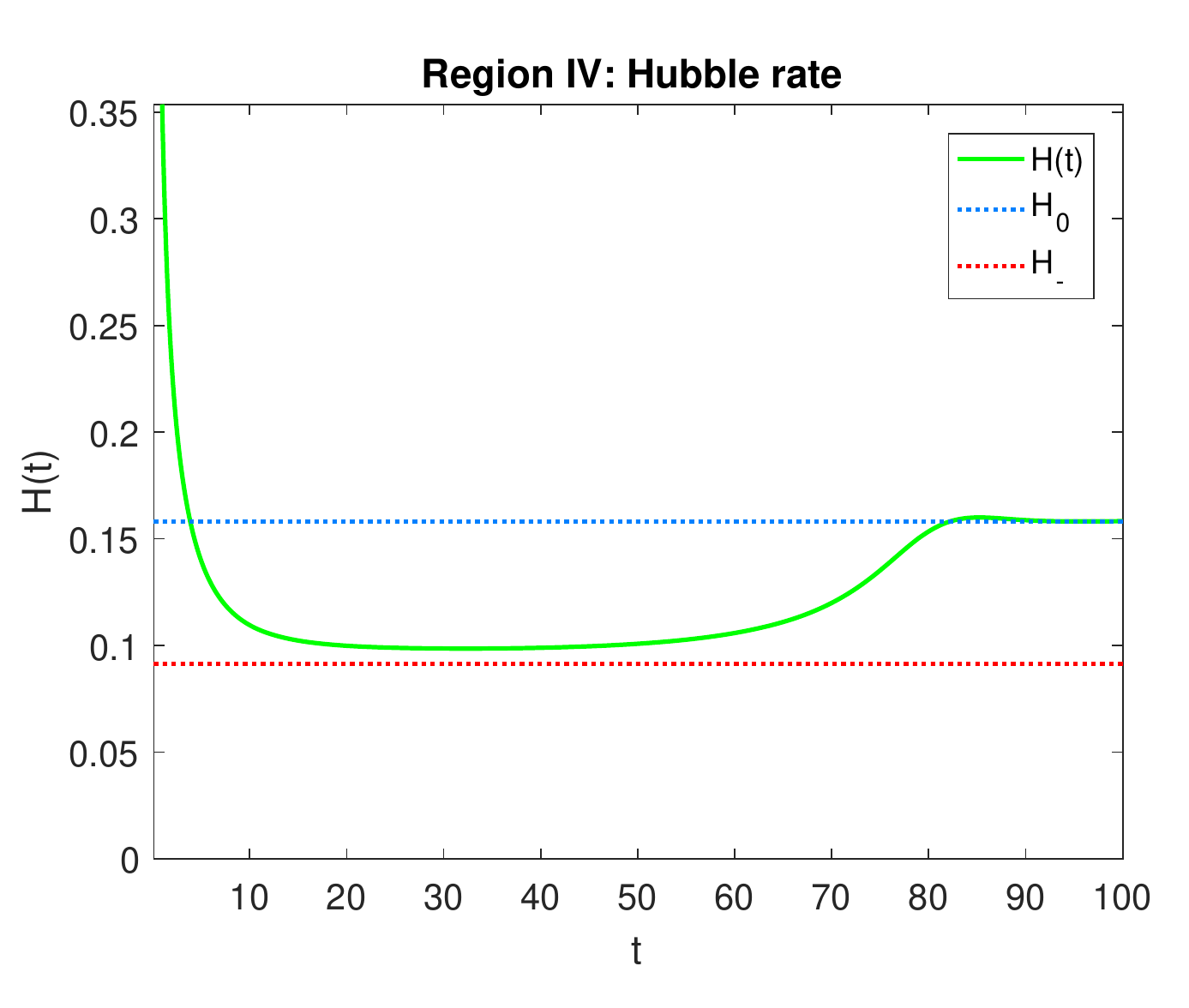}
		\caption{}
		\label{fig:regIV:3}
	\end{subfigure}
	\begin{subfigure}[b]{.45\textwidth}
		\centering
		\includegraphics[trim=3.5cm 8cm 4cm 8cm, clip=true, width=1\linewidth]{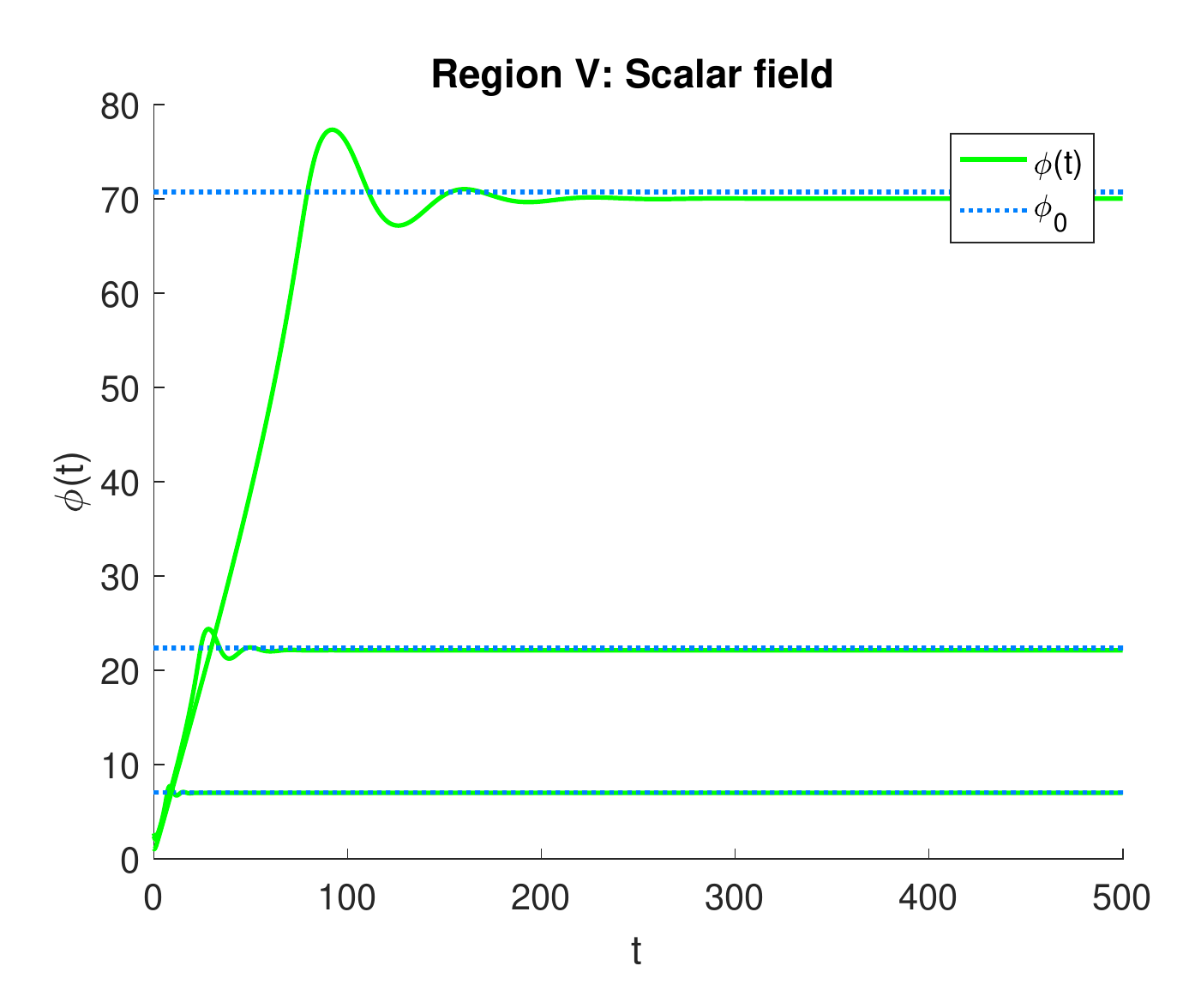}
		\caption{}
		\label{fig:regV:1}
	\end{subfigure}%
	\begin{subfigure}[b]{.45\textwidth}
		\centering
		\includegraphics[trim=3.5cm 8cm 4cm 8cm, clip=true, width=1\linewidth]{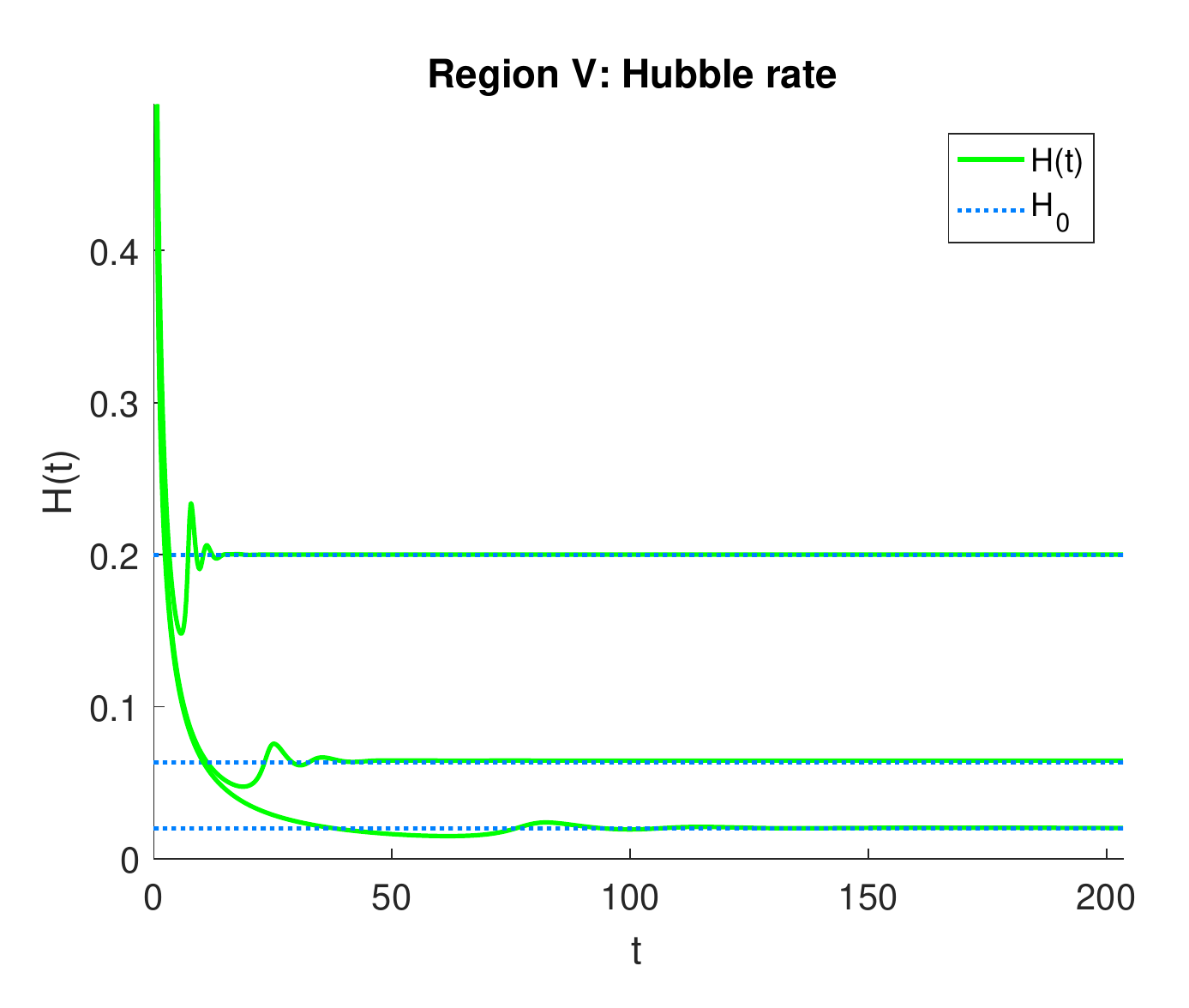}
		\caption{}
		\label{fig:regV:2}
	\end{subfigure}
	\caption{A sample of dynamical solutions (continued). Figures \ref{fig:regIII:1}-\ref{fig:regIII:2}: Region III. Figures \ref{fig:regIV:1}-\ref{fig:regIV:2}-\ref{fig:regIV:3}: Region IV. As displayed in Figure \ref{fig:regIV:3}, it is interesting to note that the Hubble rate first approaches a lower value $H_- < H_0$ before settling to the attractor constant $H_0$. This lower value $H_-=\frac{m^2}{12|\xi|}$ is the second constant solution of the equations of motion (\ref{KG}-\ref{Friedmann}) for $\lambda=0$. Figures \ref{fig:regV:1}-\ref{fig:regV:2}: Region V.}
	\label{Fig:Region345}
\end{figure}

The presence of attractor solutions has been assessed numerically. However, since at late-time $H$ and $\phi$ are constant, their expressions can be easily found analytically. In general, the equations of motion (\ref{KG}-\ref{Friedmann}) admit several constant solutions, but we note that the attractor one is always the solution with largest Hubble constant $H_0$. Table \ref{table} displays the relevant quantities that characterize constant attractor solutions. These are leading approximations, and evaluation of the residual errors should be performed on a case-by-case basis. However, as we are interested in limits where the parameters are extremely small in order to produce damping on the scale $G_{eff}H_0^2 \sim 10^{-120}$, these residual errors are insignificant. For simplicity, we have studied regions of the parameter space for which there are large separations of scales ($\ll$) between all parameters. Exceptions to this rule have been made in some cases where $\lambda, \lambda_R$ or $Gm^2,\lambda,\lambda_R$ are of the same order of magnitude ($\sim$), as we consider these cases to be of potential physical relevance. In Figures \ref{Fig:Region12} and \ref{Fig:Region345}, we display a sample of numerical solutions, all of which can be seen to approach late-time attractors. For practical and illustrative purposes, we have used moderate hierarchies of couplings in running these numerical simulations.\\

With the assumptions stated above, effective Hubble rate damping is achieved, as can be seen by reading off $G_{eff}H_0^2$ from Table \ref{table}. The damping can be arranged in many different regimes, and its strength is set by the smallness of  various dimensionless parameters: $Gm^2$, $\lambda$ or $\lambda_R$, depending on the concrete damping regime implemented. Note that damping in regions IV and V is more efficient as it is quadratic rather than linear in those parameters. The Hubble rate $H_0$ and effective Newton constant $G_{eff}$ are generically both damped, except in region I where the damping only applies to $G_{eff}$. In the appendix, we also describe non-constant attractor solutions found in the free case $\lambda=\lambda_R=0$, and in a particular subsector of Region I. For these solutions, the scalar field grows without bound, resulting in a vanishing of $G_{eff}$, as in the early works \cite{Dolgov,Ford:1987de}.\\

One might object that we have traded one small number for another, having replaced the desired small cosmological constant by a small parameter in the Lagrangian, say, $\lambda$ or $\lambda_R$. We emphasize, however, that there is a qualitative difference between having multiple known large contributions cancel each other with an incredible precision (fine-tuning) and having hierarchies among couplings in the theory, whose values are a priori undetermined. Large hierarchies, for instance, are ubiquitous in the Standard Model of elementary particles (cf. the hierarchy between the top quark and neutrino masses), and although they do call for an explanation, they are usually perceived as much less severe puzzles than fine-tuning problems. Progress is thus attained in our analysis already at this stage. We shall now take one step further and suggest a way to generate small scalar field couplings from an additional QCD-like fermion condensate, which will result in a theory whose nonzero couplings are all of order one in the fundamental Planck units, while the necessary small numbers are generated by the well-established mechanism of fermion condensation.


\section{Naturalness of couplings} \label{Section:Naturalness}

We have established in the preceding section that, within the theory (\ref{actiontotal}-\ref{phiaction}), late-time expansion with a small effective Hubble rate naturally emerges, provided that the scalar field couplings $\lambda$ and/or $\lambda_R$ are small. It is logical to ask how such small couplings could arise in a natural fashion. There are few known mechanisms to generate large hierarchies in quantum field theories. The best-established one, both theoretically and experimentally, is the generation of fermion condensates due to nonperturbative effects, as in QCD \cite{Donoghue:1992dd,Horn:2016rip}. Such condensates are nonzero expectation values of fermion bilinears $\langle \bar q q\rangle$ for a quark-like field $q$, and are expected to generically arise for fermions coupled to confining gauge theories. They are known to be phenomenologically responsible for the emergence of the observed hadron physics. The mass scale $M_q\equiv \langle \bar q q\rangle^{1/3}$ is expected to be related to the cut-off scale of the theory $M$ by an expression of the sort
\begin{equation}
\label{Mq}
M_q\sim \Lambda_{strong}= M e^{-1/\alpha(M)},
\end{equation}
where $\Lambda_{strong}$ is the analog of $\Lambda_{QCD}$ which characterizes the strong coupling energy scale of the theory, and $\alpha(M)$ is a suitable coupling renormalized at the cut-off scale $M$. In particular, \eqref{Mq} expresses the nonperturbative nature of the condensate, and may be derived by inversion of the well-known formula for the running of the coupling constant $\alpha(M)$ in non-Abelian gauge theories. Note that even for moderately small $\alpha(M)$, the ratio $M_q/M$ may encompass many orders of magnitude. Such nonperturbative mechanisms, for instance, explain why it is not unnatural to have a pion of mass $\sim 100$ MeV in the Standard Model that contains energy scales many orders of magnitude higher \cite{Donoghue:1992dd}.\\

One can easily incorporate fermion condensation in our setup to generate small scalar couplings by replacing, say, the term $G\lambda_R R \phi^4$ in (\ref{phiaction}) by
\begin{equation}
G^{5/2}\lambda_q R \phi^4 \bar q q,\label{L_q}
\end{equation}
where $\lambda_q$ is of order unity. A similar replacement can obviously be performed for the remaining small couplings. Then, since $\bar q q$ acquires a vacuum expectation value, one will effectively obtain the action (\ref{phiaction}) with 
\begin{equation}
\lambda_R=G^{3/2} \langle \bar q q\rangle = (G M_q^2)^{3/2}.
\end{equation}
This number can be as small as one wishes, provided that the fermion $q$ condenses at a scale far below the fundamental Planck scale (as quarks do in QCD), and the mechanism of cosmological constant damping from the previous section will work without any small dimensionless numbers present in the underlying Lagrangian. It could be very tempting to contemplate whether the field $q$ could be related to the actual QCD quarks and their condensates, and in terms of the scales involved, this may even appear realistic. Such attempts are however likely to run into severe phenomenological constraints in the context of hadron physics, and we shall not try to pursue this possibility here.\\

While the values of the couplings $\xi$ and $\lambda_q$ that we have introduced are natural, i.e. of order one in Planck units, and sufficient for implementing the cosmological constant damping mechanism, one may still wonder how natural is the pattern of nonvanishing couplings we have enforced. Ideally, from the standpoint of effective field theories, one should consider all possible couplings allowed by the specified set of fields and symmetries. Then, at low energies, only the terms with lowest dimensions matter. Admittedly, we do not have a setup of this form since there is no obvious symmetry assumption that would isolate the set of terms that we need for our scenario, and we do not see a way to implement such a `perfectly natural' scheme without complicating our framework considerably.\\

There is, however, a less demanding form of naturalness that our setup respects, in the spirit of \textit{softly broken symmetry}. In the context of QCD, smallness of the pion mass is in fact understood as a result of the soft breaking of chiral symmetry by quark masses\cite{Donoghue:1992dd,Horn:2016rip}. Regarding the model proposed here, we may impose the shift symmetry $\phi\to \phi + a$ on the Lagrangian \eqref{phiaction} in the UV, which would eliminate all terms except $(\partial\phi)^2$. Then, we can consider a soft breaking of this symmetry a few orders of magnitude below the Planck scale by $\xi R\phi^2$ and (\ref{L_q}). Generation of these couplings is viewed here as an entry on the wishlist for the underlying microscopic theory, whose nature is unknown to us. Consistency of this softly broken symmetry scheme then requires that the renormalization flow to lower energy scales does not upset the pattern of couplings introduced. The interaction terms considered here are perturbative below Planckian curvature and the renormalization question is therefore well-posed. Renormalization can lead to new higher-curvature terms, which are suppressed, or to insignificant changes in $\xi$, $m^2$, $\lambda$ and $\lambda_R$. Systematics of renormalization in curved spacetimes is given in \cite{Parker,Bunch:1978yq} (see \cite{Martini:2018ska} for a very recent related work on the subject).  One can convince oneself that no further couplings are generated and our construction thus passes a naturalness test at the level of a softly broken shift symmetry scheme.\\

One further naturalness-related question deserves being highlighted. Our scenarios are characterized by large contrasts between the early ($G$) and the late-time ($G_{eff}$) effective Newton's constant. The early-time value is much higher and, in particular, this drop in the value of Newton's constant contributes to the smallness of $G_{eff}H_0^2$ observed late times, in accordance with demands of realistic cosmologies. If one were to naively extrapolate the contemporary spectrum of elementary particles to such early hypothetical stages of the Universe evolution, one would end up with a situation where their masses are much higher than the early-time Planck scale. One might then worry whether the early-time Planck scale is `natural' in the sense of being protected from radiative corrections due to extremely massive particles. We emphasize that there is no sharp paradox here. First, there is no reason to extrapolate the known particle spectra to super-early cosmological stages. Second, we have no analytic control over trans-Planckian renormalization, which would necessarily involve strongly coupled non-perturbative quantum gravity. Qualitatively, even if one takes this issue seriously, one could hypothesize that some form of spacetime discreteness is present at Planckian scales, in which case trans-Planckian fields become nondynamical and freeze out, or that the degrees of freedom relevant at early cosmological stages are exactly massless and thus do not introduce any extra scales. Be it as it may, we emphasize that this potential `naturalness' issue is on a very different footing from the naturalness problem for the cosmological constant: the possible trans-Planckian naturalness issue for $G$ we have just mentioned comes from conjectural trans-Planckian physics (and thus may or may not be there), while $\Lambda$ receives many contributions from known, controllable low-energy physics, which are unacceptably large, and which are successfully diluted away in the scenarios presented here. 

\section{Summary and outlook}
We have presented a broad and simple class of scenarios that successfully realize \textit{dynamical damping of the effective cosmological constant}, an idea put forward by A. Dolgov in 1980es \cite{Dolgov}. These scenarios are based on scalar-tensor theories with negative quadratic nonminimal coupling, $\xi <0$, and are stabilized through additional quartic interactions. They possess late-time de Sitter attractor solutions whose Hubble rates are much lower than what is expected from the bare cosmological constant term of the Lagrangian, and they therefore provide a dynamical solution to the fine-tuning or ``old'' cosmological constant problem. In contrast to early attempts, these solutions do \textit{not} suffer from modifications of General Relativity at late-times, due to the stabilizing quartic interaction terms. In this way, the cosmological constant fine-tuning is traded for a \textit{hierarchy} of field couplings in our Lagrangian. This can already be viewed as a qualitative improvement by which the status of the cosmological constant problem is upgraded  from the ``worst prediction in the history of theoretical physics'' to a hierarchy puzzle of the sort commonly found in particle physics. Building on an analogy to the pion mass in QCD, we have discussed a possible way to solve this hierarchy problem based on \textit{fermion condensation} and \textit{softly broken shift symmetry}. This results in an attractive realization of the cosmological constant damping starting with a Lagrangian that only contains couplings of order one in Planck units (explaining the specific set of nonvanishing couplings we have retained to generate the necessary cosmological solutions remains at this stage a requirement for the underlying microscopic theory). \\

Many directions for future investigations can be suggested. We have shown strong numerical evidence that the proposed class of models possesses an \textit{attractor behavior}, however a rigorous mathematical proof of this fact is still lacking. In addition, if such a model is to solve the old cosmological constant problem definitively, many more phenomenological constraints have to be accommodated. For this, it should be embedded in the standard $\Lambda$CDM model of cosmology, and its consequences on the evolution of the Universe should be investigated. In order to preserve the numerous successes of modern cosmology, one naturally infers that the late-time attractor solution has to settle in well before recombination and galaxy formation. Two main challenges then arise: to accommodate radiation- and matter-dominated eras, as well as the theory of inflation. See \cite{Barr:2006mp} for a similar discussion in an earlier attempt at dynamical damping of the cosmological constant. We should emphasize that the \textit{simplicity} and \textit{genericity} of the class of models proposed here allows for adaptation or embedding into more elaborate scenarios in many different ways, either from a theoretical or from a phenomenological perspective. Such robustness can be seen as a reward for having insisted on working with a consistent effective field theory.

\section*{Acknowledgments}
We thank Emil Akhmedov, Ignatios Antoniadis, Chethan Krishnan, Roman Pasechnik and Richard Woodard for related discussions, and Ott Vilson for drawing our attention to a few algebraic typos. We have also benefitted from correspondence with Frans Klinkhamer and Viacheslav Emelyanov on issues of naturalness. The work of OE is funded by CUniverse research promotion project (CUAASC) at Chulalongkorn University. The work of KN is supported in part by FWO-Vlaanderen through project G006918N and by Vrije Universiteit Brussel through the Strategic Research Program ``High-Energy Physics.'' OE is grateful to the physics department of the Jagiellonian University in Krak\'ow, and especially to Piotr Bizo\'n, for hospitality during the concluding stages of this work.

\appendix
\section*{Appendix: Non-constant attractor solutions} 

We summarize our understanding of \textit{non-constant} attractor solutions of the free massive scalar field ($\lambda=\lambda_R=0$) and of a particular subsector of Region I ($Gm^2, \lambda_R \ll \lambda$).

\subsection*{Free massive scalar field}
The free massless scalar  has been studied by Dolgov \cite{Dolgov} and Ford \cite{Ford:1987de}. The free massive scalar field has growing solutions if the following condition is satisfied:  
\begin{equation}
\label{growingsolution}
m^2+12\xi H_\Lambda^2<0.
\end{equation}
In this case there is a late-time attractor behavior,
\begin{align}
\phi(t) &= \phi_0\ e^{(\nu_0-3/2)H_0 t}+\phi_{sub}\ e^{-(\nu_0-3/2)H_0 t}+...,\\
\label{late-time}
H(t) &= H_0+H_{sub}\ e^{-(2\nu_0-3)H_0 t}+...,
\end{align}
with
\begin{align}
\nu_0&=\frac{3-16\xi}{2-8\xi}>\frac{3}{2},\\
H_0^2&=\frac{m^2}{9/4-12\xi-\nu_0^2},\\
H_{sub}&=\frac{1}{|\xi|(1-6\xi)G\phi_0^2}\frac{H_\Lambda^2-H_0^2}{H_0},\\
\phi_{sub}&=-\frac{3(1-4\xi)}{2}\frac{H_{sub}}{H_0}\phi_0.
\end{align}
Here, $\phi_0$ is undetermined from the late-time analysis and depends on the initial conditions through the detailed dynamics of the system. In this sense, the attractor solution is not unique, although its behavior is universally described by the above equations. The situation here is similar to the massless scalar field \cite{Dolgov,Ford:1987de}: although the Hubble rate approaches a constant exponentially fast, $\phi$ does not stabilize and the effective Newton constant is therefore time-dependent and asymptotically vanishing. 

\subsection*{Special cases in Region I}
The attractor solution of Region I described in Section \ref{Section:Attractor} is found by taking the limit $\lambda_R\to 0$.\footnote{One can check that the limits $m^2\to 0$ and $\lambda_R \to 0$ commute, such that this solution also describes the limit $Gm^2\ll \lambda_R \ll 1$.} In this case, the equations of motion admit a unique constant solution,
\begin{align}
\label{phiI}
\phi_0^2=\frac{12 \xi H_\Lambda^2-m^2}{2\lambda+\xi Gm^2}, \qquad H_0^2=\frac{48\lambda H_\Lambda^2-Gm^4}{48\lambda+12\xi Gm^2}. 
\end{align}
In the limit $Gm^2 \ll \lambda \ll 1$, one recovers the expressions of Table \ref{table}. If the expression \eqref{phiI} for $\phi_0^2$ is negative, then this solution does not exist and there is no constant attractor solution. Instead, one finds a \textit{linearly growing} attractor behavior at late-time,
\begin{align}
\phi(t)&=\phi_1\ t^{1/3} + \phi_0 +\mathcal{O}(t^{-1/3}),\\
H(t)&=H_1\ t^{1/3} + H_0 +\mathcal{O}(t^{-1/3}),
\end{align}
satisfying the relations
\begin{equation}
\phi_1^2=-\frac{6\xi}{\lambda} H_1^2, \qquad \phi_0^2=-\frac{6\xi}{\lambda} H_0^2,
\end{equation}
such that
\begin{equation}
G_{eff} H^2= \frac{\lambda}{3\xi^2} +\mathcal{O}(t^{-1/3}).
\end{equation}
Although the effective Hubble rate is constant, the solution suffers from vanishing of the gravitational strength, $G_{eff} \sim \phi^{-2} \sim t^{-2/3}$.

\end{document}